\documentclass[nofootinbib,11pt,showpacs]{revtex4-1}
\usepackage{graphics}
\usepackage{mathptmx}
\usepackage[dvips]{graphicx}
\usepackage{mathrsfs}
\usepackage{amssymb}
\usepackage{amsmath}
\usepackage{cancel}
\usepackage{verbatim}
\usepackage{float}
\usepackage{slashed}
\usepackage{soul}
\usepackage{xcolor}
\usepackage{bbm}
\usepackage{ulem}
\usepackage{setspace}
\usepackage[dvips,letterpaper,text={6.5in,9in}]{geometry}
\usepackage{paralist}

\usepackage{tikz}
\usetikzlibrary{trees}
\usetikzlibrary{positioning,arrows}
\usetikzlibrary{decorations.pathmorphing}
\usetikzlibrary{decorations.markings}
\definecolor{jblue}  {RGB}{20,50,100}
\definecolor{npurple}  {RGB} {153, 51, 204}
\definecolor{wred}   {RGB}{217,0,56}
\definecolor{white}   {RGB}{255,255,255}

\definecolor{korange}   {RGB}{235, 80,  43}
\definecolor{korange2}   {RGB}{245, 100,  63}
\definecolor{kyelloworange}   {RGB}{255, 210,  110}
\definecolor{kyelloworange2}   {RGB}{240, 170,  90}
\definecolor{kred}   {RGB}{204,  102, 153}
\definecolor{kpurple}   {RGB}{153,  61, 190}
\definecolor{kpurplelight}   {RGB}{213,  161, 230}

\tikzset{
	photon/.style={decorate, decoration={snake}, draw=black,very thick},
	boson/.style={decorate, decoration={snake}, draw=black,very thick},
	electron/.style={draw=black,very thick, postaction={decorate},
		decoration={markings,mark=at position .55 with {\arrow[draw=jblue]{>}}}
	},
	electron2/.style={draw=black,very thick, postaction={decorate},
		decoration={markings,mark=at position .55 with {\arrow[draw=jblue]{<}}}
	},
	fermion/.style={draw=jblue,very thick, postaction={decorate},
		decoration={markings,mark=at position .55 with {\arrow[draw=jblue]{}}}
	},
	gluon/.style={decorate, draw=black,very thick, 
		decoration={coil,amplitude=4pt, segment length=6pt}},
	higgs/.style={draw=wred,very thick, postaction={decorate},
		decoration={markings,mark=at position .55 with {\arrow[draw=wred]{>}}}
	},
	nothing/.style={draw=white,very thick}
}

\usepackage[english]{babel}
\usepackage{hyperref}
\hypersetup{
	colorlinks=true,
	linkcolor=blue,
	filecolor=magenta,
	urlcolor=cyan,
}
\usepackage{tabu, multirow,color,dcolumn,bm,enumerate}

\usepackage{graphicx,bm}

\newcommand{\beq}{\begin{equation}}
	\newcommand{\bea}{\begin{eqnarray}}
		\newcommand{\eeq}{\end{equation}}
	\newcommand{\eea}{\end{eqnarray}}
\newcommand{\bal}{\begin{align}}
	\newcommand{\eal}{\end{align}}

\usepackage{latexsym}
\usepackage{subcaption}

\usepackage{outlines}
\usepackage{enumitem}
\setenumerate[1]{label=\Roman*.}
\setenumerate[2]{label=\Alph*.}
\setenumerate[3]{label=\roman*.}
\setenumerate[4]{label=\alph*.}
\usepackage[compact]{titlesec}
\titlespacing{\section}{0pt}{2ex}{1ex}
\titlespacing{\subsection}{0pt}{1ex}{0ex}
\titlespacing{\subsubsection}{0pt}{0.5ex}{0ex}

\definecolor{darklightsabergreen}{rgb}{0.0, .49, 0.06}

\usepackage[font=small,labelfont=bf,justification=justified]{caption}
\usepackage{ragged2e}

\usepackage{graphicx}
\usepackage{float}
\usepackage{dcolumn}
\usepackage{bm}

\usepackage{epsfig}
\input epsf.tex
\usepackage{amsthm}
\usepackage{amsopn}

\def\comment#1{}

\def\beq{\begin{equation}}
	\def\eeq{\end{equation}}
\def\bea{\begin{eqnarray}}
	\def\eea{\end{eqnarray}}

\def\ep{\epsilon\!\!\!/}

\usepackage{ulem}
\usepackage{cancel}
\usepackage{color}

\begin{document}
	{\textsf{\today}
		\title{
			Cross-correlation Power Spectra and Cosmic Birefringence of the CMB via Photon-neutrino Interaction
		}
		\author{Roohollah Mohammadi}
		\email{rmohammadi@ipm.ir}
		\affiliation{ Iranian National Museum of Science and Technology (INMOST), PO Box 11369-14611, Tehran, Iran.\\
			School of Astronomy, Institute for Research in Fundamental
			Sciences (IPM), P. O. Box 19395-5531, Tehran, Iran.
		}
		\author{Jafar Khodagholizadeh}
		\email{gholizadeh@ipm.ir}
		\affiliation{%
			Farhangian University, P.O. Box 11876-13311,  Tehran, Iran.
		}%
		\author{Mahdi Sadegh}
		\email{m.sadegh@ipm.ir}
		\affiliation{
			School of Particles and Accelerators, Institute for Research in Fundamental
			Sciences (IPM), P. O. Box 19395-5531, Tehran, Iran.
		}
		%
		\author{Ali Vahedi}
		\email{vahedi@ipm.ir}
		\affiliation{Department of Astronomy and High Energy Physics,\\ Faculty of Physics, Kharazmi University, P.~O.~Box 15614, Tehran, Iran.
		}
		\author{S. S. Xue}
		\email{xue@icra.it}
		\affiliation{ ICRANet Piazzale della Repubblica, 10 -65122, Pescara, Italy.\\
  Physics Department, University of Rome La Sapienza, Rome, Italy.\\
   INFN, Sezione di Perugia, Perugia, Italy\\
    ICTP-AP, University of Chinese Academy of Sciences, Beijing, China.}
		
		\begin{abstract}
			In the context of the standard model of particles, the weak interaction of cosmic
			microwave background (CMB) and cosmic neutrino background (C$\nu$B), can generate non-vanishing TB and EB power spectra in the order of one loop forward scattering, in the presence of scalar perturbation, which is in contrast
			with the standard scenario cosmology.
		Comparing our results with the current experimental data may provide, significant information about the nature of C$\nu$B, including CMB-C$\nu$B forward scattering for TB, TE, and EB power spectra. To this end, different cases were studied, including Majorana C$\nu$B and Dirac C$\nu$B. On the other hand, it was shown that the mean opacity due to cosmic neutrino background could  behave as an anisotropic  birefringent medium and change the linear polarization rotation angle. Considering the contributions from neutrino and anti-neutrino forward scattering with CMB photons (in the case of Dirac neutrino), we introduce relative neutrino and anti-neutrino density asymmetry ($\delta_\nu=\frac{\Delta n_\nu}{n_\nu}=\frac{n_\nu-n_{\bar{\nu}}}{n_\nu}$). Then, using the cosmic birefringence angle reported by the Planck data release $\beta=0.30^\circ\pm0.11^\circ$ ($68\%C.L.$), some constraints can be put on $\delta_\nu$. Also, the value of cosmic birefringence due to Majorana C$\nu$B  medium is estimated at about $\beta|_\nu\simeq0.2$ rad. In this respect, since Majorana neutrino and anti-neutrino are exactly the same, both CB contributions will be added together. However, this value is at least two orders larger than the cosmic birefringence angle reported by the Planck data release,  $\beta=0.30^\circ\pm0.11^\circ$ ($68\%C.L.$). Finally, we shortly discussed this big inconsistency. It is noteworthy that to calculate the contribution of photon-neutrino forward scattering for cosmic birefringence, we just consider the standard model of particles and the standard scenario of cosmology. 
			
		\end{abstract}
		\keywords{Tensor modes fluctuations, Curved cosmology, Neutrinos.}
		\pacs{13.15.+g,34.50.Rk,13.88.+e}
		
		\maketitle
		
		\section{Introduction}
		Cosmic neutrinos can decouple deep earlier from matter about 1 sec after the Big Bang (BB) at the temperature of $\cong 1 MeV \cong 10^{10} K$. Therefore, along with the cosmic microwave background (CMB) coming from 380000 years after BB, the cosmic neutrino background (C$\nu$B) contains valuable information about the history of the universe \cite{weinberg}. Today, It is very hard to detect C$\nu$B due to its very weak interactions and low energy ($1.95 K\cong 1.68\times 10^{-4}eV$).\\
		Moreover, a relativistic neutrino beam can not help make a difference, at least in observation, between the Dirac and Majorana neutrino in the case of electromagnetic interaction. This fact directly relates to the left-handed interacting neutrino in the SM context, which is called the ”Practical Dirac-Majorana confusion theorem” \cite{kayser}. However the very low energy of C$\nu$B that makes these particles is in the  non-relativistic regime. Therefore, the Dirac or Majorana nature of neutrino has different electromagnetic effects \cite{palbook} detected by upgrading experiments. Recently, some experiments have been conducted to detect C$\nu$B. For example, KATRIN equipped to C$\nu$B by the most probable interaction channel to capture the electron neutrino from the background by Tritium ($\nu_e(1.95 Kelvin)+{}^{3}H\rightarrow {}^{3}He+e^-$) \cite{Faessler:2016tjf}.
		Nevertheless, many theoretical attempts have been proposed in the literature to search for these relic neutrinos \cite{Faessler:2011qj,Lazauskas:2007da,Munyaneza:2006kw,Pas:2001nd,weiler,Wigmans:2002rb,Ringwald:2003qa,Ringwald:2004np}.
		The detection method of C$\nu$B is divided into three categories: (1) direct detection via momentum transfer in neutrino elastic scattering off nuclei target, (2) neutrino capture in nuclei, or (3) indirect method by spectral distortion generated by C$\nu$B and ultra-high energy neutrinos or protons originating from unknown sources \cite{yanagisawa}.  The common issue of all these categories is that no search has been done in any of them for the nature of Dirac or Majorana neutrinos.
		Another indirect detection method of C$\nu$B  was proposed by Mohammadi et al. \cite{Mohammadi:2016bxl,Khodagholizadeh:2014nfa,Mohammadi:2013dea,Mohammadi:2013ksa}. In this method, photon-neutrino forward scattering is considered a tool  for probing C$\nu$B that distinguishes the Majorana and Dirac nature of C$\nu$B.\\
		The CMB photons can interact electromagnetically with matter. Therefore, it is possible to test very vast theoretical models of particle physics and  cosmology \cite{Abazajian:2013oma,Abazajian:2016yjj,Jeong:2019zaz,Kamionkowski:2015yta,Zucca:2016iur,Kosowsky:1994cy}. The intensity and polarization of CMB are used to extract the required information.\\
		The most commonly used  observables to investigate theoretically and experimentally are the CMB polarization power spectra. The B-mode ($C_\ell^{BB}$) power spectrum can be used to study primordial gravitational waves \cite{Seljak:1996gy}.  In the standard scenario of cosmology, the CMB B-mode polarization due to the Thomson scattering can only be generated in the presence of tensor metric perturbations \cite{Zaldarriaga:1996xe, Zaldarriaga:1997ch, Hu}. Gravitational lensing is also another source of B-mod \cite{SPTpol:2013omd}.  Thereupon, the B-mode power spectrum can calculate one  of the most important cosmological parameters as tensor-to-scalar (r-parameter) \cite{cardoso,Vazquez:2013dva}. E-mode polarization is generated by both scalar and tensor metric fluctuations. Nowadays, E-mode power spectrum ($C_\ell^{EE}$) measurements are almost consistent with the $\Lambda$CDM cosmological model \cite{Henning:2017nuy,Aghanim:2019ame,Louis:2016ahn}. It is of note that temperature-polarization power spectrum ($C_\ell^{TE}$) measurements can surpass our information about CMB temperature anisotropy ($C_\ell^{TT}$) and put some stronger constraints on cosmological model. This effect is attributed to the lower levels of polarization foreground measurements \cite{Galli,Louis:2016ahn}. This point motivates us to investigate some of the main cross-correlation power spectra such as $TE$, $TB$, and $EB$.\\
		In the standard scenario of cosmology, cross-correlations of $EB$ and $TB$ are due to the parity invariant  vanishing \cite{Kamionkowski:1996zd,Seljak:1996gy}. Thus,  these cross-correlations can be used  to check some parity-violating interactions and go beyond the standard models \cite{kamion1999}. In \cite{liu,Carroll}, the authors have theoretically investigated the quintessence and Axion-like dark matter particle coupled to the electromagnetic strength tensor and it's dual ($\mathcal{L}=\phi(t)F_{\mu\nu}\tilde{F}^{\mu\nu}$). Besides,  they showed that can generate non-zero $EB$ and $TB$ power spectra. In these theoretical models,   the plane of CMB linear polarization (E-mode and B-mode polarizations)  can rotate by an angle proportional to $\int^{t_0}_{t_{\rm LSS}} dt~\dot{\phi}$  which  is known as cosmic birefringence (CB) \cite{Gluscevic}. In another paper \cite{Geng} authors proposed a CPT-even six-dimensional effective Lagrangian to consider the effect of neutrino density asymmetry on CB angle. The authors in \cite{Geng} did not discuss the nature of neutrinos in their calculation. But, in this work, we combine the result of \cite{Geng} on the density asymmetry of neutrino and anti-neutrino and the direct one-loop calculation of neutrino (anti-neutrino)-photon scattering amplitude for Dirac neutrino. In addition, we calculate the same method separately for Majorana neutrino. In conventional physics, the cosmological birefringent medium is caused by some radiation anisotropies of the background along the cosmological path of CMB since the last scattering surface time. These anisotropies, for example, include finite displacement of electrons in the ionized plasma, photon-photon scattering due to the virtual electron-positron pairs in a vacuum, and lack of hydrogen atom potential as a simple harmonic oscillator  \cite{Montero-Camacho:2018vgs}.\\  
		Moreover, in the presence of cosmic neutrino background, their interaction with cosmic microwave background via parity violating forward scattering makes an opaque medium: this opacity acts as a birefringent medium. Using the phenomenological value of this opacity, we can calculate the contribution of CMB-C$\nu$B interaction at the measured CB angle. Recently, This CB has been measured by analyzing the 2022 cosmological foreground polarization data of Planck ($\beta=0.3^\circ\pm0.11^\circ$) for small $l$ at a 68\% confidence level \cite{minami}. Also, some other experiments have measured CB that rely on polarization angle calibration based on hardware calibrators and astrophysical sources \cite{Ade:2014afa,Bianchini:2020osu,Namikawa:2020ffr}. Contrary to \cite{minami} their results were consistent with zero. \\
		In \cite{Mohammadi:2016bxl},
		we studied the generation of CMB B-mode polarization via forward scattering of photon-neutrino in the case of tensor metric perturbation. The result showed that the effect of Majorana neutrinos on the $r$-parameter was different from that of the Dirac neutrinos.
		\\ Distinguishing Dirac from Majorana neutrinos can be used in the microwave cavity \cite{Abdi:2019hvx},
		$SU(6) $ grand unified theories \cite{Chacko:2020tbu}, weak interaction of aggregate matter \cite{Segarra:2020rah}, neutrino decay\cite{Balantekin:2018ukw}, pure leptonic decays at the LHC \cite{Arbelaez:2017zqq}, neutrino-electron elastic scattering in the presence of all possible Lorentz-invariant interactions \cite{Rodejohann:2017vup}, and non-cyclic geometric phase for neutrinos \cite{Capolupo:2016idi,Lu:2021yav,Johns:2021ets}.\\
		In this work, we will solve the quantum Boltzmann equation for the time-evolution of matrix density (Stokes parameters) of an ensemble of CMB photons.  The CMB-C$\nu$B scattering is involved here as a collision term. The CMB-C$\nu$B weak interaction not only can generate non-zero EB and TB power spectra but also can identify the Dirac and Majorana nature of C$\nu$B. Moreover, the generated neutrino opacity via CMB-C$\nu$B interaction behaves as a birefringent medium that can change the CMB linear polarization angle. Therefore, the observed cross-power spectra between E-mode, B-mode, and temperature in the background of scalar perturbations have a significant contribution to measured CB. Hence, this scattering should be regarded among the most effective phenomena. 
		To give an effective sense, we show that the CB angle $\beta$ can be roughly estimated by the value of opacity in $z=1100$ due to Majorana C$\nu$B  medium  $\beta|_\nu\simeq1/2\kappa(z=1100)$ where $\kappa=\int_\eta^{\eta_0} d\eta \dfrac{\sqrt{2}}{6\pi} \alpha G^{F}\dfrac{n_{\nu}}{2} $, $\eta$ is the conformal time, $n_{\nu}$ is neutrino number density, $G^{F}$ and $ \alpha $ are Fermi coupling constant and electromagnetic fine structure constant. The estimated CB angle in the case C$\nu$B as Majorana particle is $\beta|_\nu\simeq 0.2~rad$ (see Section III for details). However this value is at least two orders larger than the cosmic birefringence angle reported by the Planck data release, $\beta\simeq0.0053$rad ($68\%C.L.$). Thus, let us see what can be the source of this big inconsistency. First, the electromagnetic interaction of CMB and C$\nu$B is not the only source of CB angle. To better understand the mentioned inconsistency, all other possible sources for CB should be considered. We have shortly discussed this big inconsistency before the conclusion. 
		\section{Cross power spectrum in the presence of Thomson  and CMB-C$\nu$B scattering}
		To calculate the cross-power spectra of CMB radiation, we start with CMB radiation as an ensemble of photons described by
		\begin{eqnarray}\label{rho}
			&&\hat{\rho}_{ij}=\dfrac{1}{tr(\hat{\rho})}\int \dfrac{d^{3}k}{(2\pi)^{3}}\rho_{ij}(k)D_{ij}(k),\nonumber\\  &&\hat{\rho}_{ij}\equiv\frac{1}{2}\left(
			\begin{matrix}
				I+Q & U-iV \\
				U+iV&  I-Q \\
			\end{matrix}
			\right)
		\end{eqnarray}
		where $ I $, $Q$ ($ U $) and $V$ are the Stokes parameters indicating the total intensities, linear polarizations intensities and the difference between clockwise and counter-clockwise circular polarization intensities, respectively \cite{Chandra}. $ D_{ij} (\vec{k})\equiv a_{i}^{\dagger}(\vec{k})  a_{j}(\vec{k}) $ is the photon number operator.
		Then, to obtain the time evolution of Stokes parameters, we solved the following quantum Boltzmann equation as follows:
		\begin{eqnarray}
			(2\pi)^32k^0 \frac{d}{dt}\rho_{ij}(\boldmath k)=i<[H_I^0,D_{ij}(\boldmath k)]>
			-\frac{1}{2}\int_{-\infty}^{+\infty}dt<[H_I^0(t),[H_I^0(t),D_{ij}(\boldmath k)]]>,
			\label{QBE}
		\end{eqnarray}
		where  $H_I^0$ denotes the leading order term of the interacting Hamiltonian. The first term on RHS of the above equation is called forward scattering (interacting without momentum exchange) with a linear dependence on $H_I^0$. The second one is the high-order collision terms (representing ordinary cross-section terms) with square dependence on $H_I^0$ (see \cite{Kosowsky:1994cy} and Appendix for more detail). As shown in \cite{Mohammadi:2013ksa}, in the case of the C$\nu$B-CMB interaction given by Fig.(\ref{fig:photon-neutrino}), the contribution of forward scattering is non-zero and proportional to ($\propto\alpha G_F$). Note that, the usual cross-section of photon- neutrino scattering is proportional to ($\propto\alpha^2G_F^2$), which is very smaller than the forward scattering term.
		\begin{figure}
			\begin{center}
				\begin{tikzpicture}[scale=0.90,
					thick,
					level/.style={level distance=3.15cm, line width=0.4mm},
					level 2/.style={sibling angle=60},
					level 3/.style={sibling angle=60},
					level 4/.style={level distance=1.4cm, sibling angle=60}
					]
					\node[draw=none,fill=none] (p1) at (-4,3){}  ;
					\node[draw=none,fill=none] (v1) at (-2,2){}  ;
					\node[draw=none,fill=none] (v2) at (0,2){}  ;
					\node[draw=none,fill=none] (v3) at (-2,0.0){}  ;
					\node[draw=none,fill=none] (v4) at (0,0){}  ;
					\node[draw=none,fill=none] (p2) at (-4,-1.5){}  ;
					\node[draw=none,fill=none] (k1) at (1.7,3) {};
					\node[draw=none,fill=none] (k2) at (1.7,-1.5) {};
					\node[draw=none,fill=none] at (-3,3){$\gamma$}  ;
					\node[draw=none,fill=none] at (-1,-.5){$W^+$};
					\node[draw=none,fill=none] at (-1,2.5){$e^-$};
					\node[draw=none,fill=none] at (0.6,1){$e^-$};
					\node[draw=none,fill=none] at (-2.5,1){$e^-$};
					\node[draw=none,fill=none] at (-3,-0.2){$\nu_e$}  ;
					\node[draw=none,fill=none] at (0.5,3) {$ \gamma $};
					\node[draw=none,fill=none] at (1,-.2) {$ \nu_e$};
					\draw[photon]   (p1) -- (v1);
					\draw[electron] (v1) -- (v2);
					\draw[photon]   (v2) -- (k1);
					\draw[electron] (v4) -- (v3);
					\draw[electron] (v3) -- (v1);
					\draw[electron] (v2) -- (v4);
					\draw[electron] (v4) -- (k2);
					\draw[electron] (p2) -- (v3);
					
					\draw[black]         (-2,2) circle (0.1cm);
					\fill[black]    (0,0) circle (0.1cm);
					\fill[black]    (-2,0) circle (0.1cm);
					\draw[black]         (0,2) circle (0.1cm);
				\end{tikzpicture}
				\begin{tikzpicture}[scale=0.90,
					thick,
					level/.style={level distance=3.15cm, line width=0.4mm},
					level 2/.style={sibling angle=60},
					level 3/.style={sibling angle=60},
					level 4/.style={level distance=1.4cm, sibling angle=60}
					]
					\node[draw=none,fill=none] (p1) at (-4,3){}  ;
					\node[draw=none,fill=none] (v1) at (-2,2){}  ;
					\node[draw=none,fill=none] (v2) at (0,2){}  ;
					\node[draw=none,fill=none] (v3) at (-2,0.0){}  ;
					\node[draw=none,fill=none] (v4) at (0,0){}  ;
					\node[draw=none,fill=none] (p2) at (-4,-1.5){}  ;
					\node[draw=none,fill=none] (k1) at (1.7,3) {};
					\node[draw=none,fill=none] (k2) at (1.7,-1.5) {};
					\node[draw=none,fill=none] at (-3,3){$\gamma$}  ;
					\node[draw=none,fill=none] at (-1,-.5){$W^+$};
					\node[draw=none,fill=none] at (-1,1.7){$e^-$};
					\node[draw=none,fill=none] at (0.6,1){$e^-$};
					\node[draw=none,fill=none] at (-2.5,1){$e^-$};
					\node[draw=none,fill=none] at (-3,-0.2){$\nu_e$}  ;
					\node[draw=none,fill=none] at (0.5,3) {$ \gamma $};
					\node[draw=none,fill=none] at (1,-.2) {$ \nu_e$};
					\draw[photon]   (p1) -- (v2);
					\draw[electron] (v1) -- (v2);
					\draw[photon]   (v1) -- (k1);
					\draw[electron] (v4) -- (v3);
					\draw[electron] (v3) -- (v1);
					\draw[electron] (v2) -- (v4);
					\draw[electron] (v4) -- (k2);
					\draw[electron] (p2) -- (v3);
					
					\draw[black]         (-2,2) circle (0.1cm);
					\fill[black]    (0,0) circle (0.1cm);
					\fill[black]    (-2,0) circle (0.1cm);
					\draw[black]         (0,2) circle (0.1cm);
				\end{tikzpicture}
				\caption{Two Feynman diagrams describe the interaction between photons and Dirac neutrinos at the one-loop level. Along the loop, there are two QED-vertexes and two weak interacting vertexes. Note that the neutral channel diagram with a $Z^0$-boson exchange does not contribute to the interacting Hamiltonian $H^0_I$. The "conjugated" Feynman diagrams should be considered in the case of neutrinos as Majorana particles. This diagrams can be extracted by replacing all the internal lines in their conjugated
					lines. It is noteworthy that there are other diagrams containing more W bosons lines. These diagrams give contributions in the order of $\mathcal{O}(G_F^2)$. Regarding the very small contribution of these diagrams, we ignore them. }
				\captionsetup[figure]{justification=justified,singlelinecheck=true}
				\label{fig:photon-neutrino}
			\end{center}
		\end{figure}
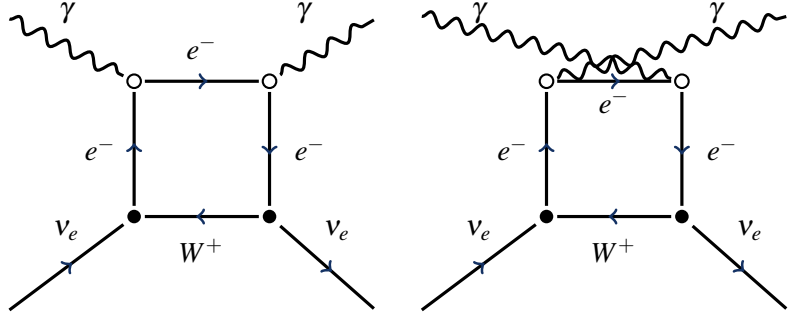
		The Boltzmann equation for the linear polarization and temperature anisotropy parameters of CMB radiation in the presence of Compton scattering  and photon-neutrino forward scattering could be written as\cite{roh2}
		{\small \begin{eqnarray}
				&&\frac{d}{d\eta}\Delta_T^{(S)} +iK\mu\Delta_T^{(S)}+4[\dot{\psi}-iK\mu \varphi]\nonumber\\
				&&~~~~~~=\dot\tau_{e\gamma}\Big[-\Delta_T^{(S)} +
				\Delta_{T}^{0(S)} +i\mu v_b +{1\over 2}P_2(\mu)\,\Pi\Big],\label{Boltzmann}\\
				&&\frac{d}{d\eta}\Delta _{P}^{\pm (S)} +iK\mu \Delta _{P}^{\pm (S)} \nonumber\\
				&&~~= \dot\tau_{e\gamma}\Big[-\Delta_{P}^{\pm (S)}-{1\over 2} [1-P_2(\mu)]\, \Pi\Big] \mp a(\eta) \dot{\kappa}~\Delta _{P}^{\pm (S)},
				\label{Boltzmann1}
		\end{eqnarray}}
		where $\Delta _{P}^{\pm (S)}=Q^{(S)}\pm i U^{(S)}$, $\Pi\equiv \Delta_{I}^{2(S)}+\Delta_{P}^{2(S)}+\Delta_{P}^{0(S)}$, $\psi$, and $\phi$ are scalar functions defining scalar perturbation of metric in longitudinal gauge,  $v_b$ is neutrino bulk velocity,  $ \dot{\tau}_{e\gamma}=a n_{e} x_{e} \sigma_{T} $ is differential optical depth for Thomson scattering, where $ a(\eta) $ is the expansion factor normalized to unity; $ x_{e} $  is the ionization fraction and $ \sigma_{T} $ is Thomson cross-section.  $\dot{\kappa}=d\kappa(\eta)/d\eta $ is opacity due to this interaction. In the case of a Majorana neutrino, this parameter has different behavior compared to the Dirac neutrino case.\\
		The CMB polarization $ \Delta_{p}^{\pm (S)} (\eta_{0},k,\mu)$ is separated into the divergence-free part (B-mode) and curl-free (E-mode) as follows:
		\begin{figure*}
			\centering
			\centering
			\includegraphics[width=0.7\linewidth]{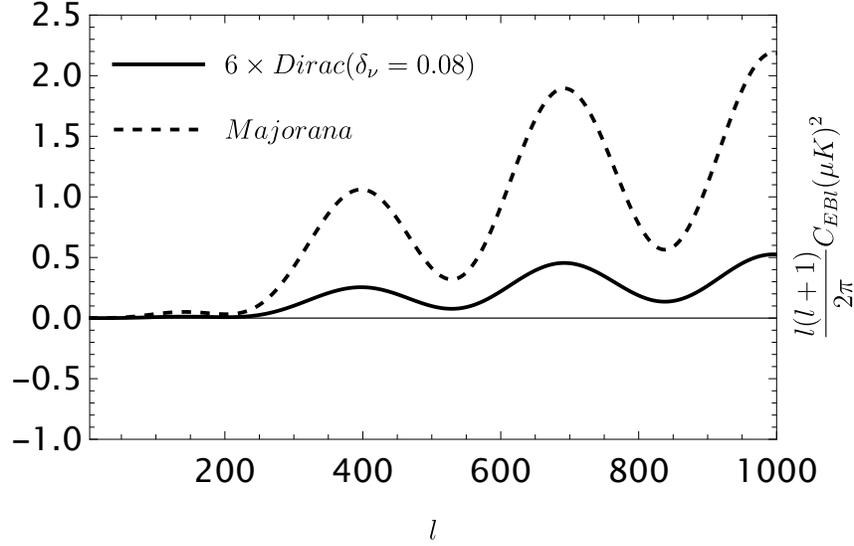}
			\caption{EB power spectrum in the presence of scalar perturbation via C$\nu$B-CMB forward scattering is plotted in terms of $l$ by considering C$\nu$B as Dirac for $\delta_\nu=0.08$ (solid line) and Majorana (dashed line) neutrino. Note in standard scenario of cosmology,  EB power spectrum in the presence of scalar perturbation is exactly zero. }
			\label{EBdata-2}
		\end{figure*}
		\begin{widetext}
			\begin{eqnarray}
				\Delta_{E}^{(S)}(\eta_0,K,\mu)&=&
				-\frac{3}{4}\int_{0}^{\eta_{0}}d\eta \,g(\eta)\,\Pi(\eta,K)\{\partial_{\mu}^{2} \left[(1-\mu^{2})^2e^{ix\mu} \cos{\kappa(\eta)}\right]\}, \label{Emode1}\\
				\Delta_{B}^{(S)}(\eta_0,K,\mu)&=&\frac{3}{4}\int_{0}^{\eta_{0}}d\eta \,g(\eta)\,\Pi(\eta,K)\{\partial_{\mu}^{2} \left[ (1-\mu^{2})^2e^{ix\mu} \sin{\kappa(\eta)}\right]\},\label{Bmode1}
			\end{eqnarray}
		\end{widetext}
		where $g(\eta)=\dot{\tau}_{e\gamma}e^{-\tau_{e\gamma}}$,  $\partial_{\mu}=\partial/\partial\mu$, and $ \mu= \hat{\bf n}.\hat{\mathbf{K}}$  is the cosine angle of the CMB photon direction  $\hat{\bf n}={\bf k}/|{\bf k}|$ and the wave vectors $\bf{K}$, $x=\bf K(\eta_0-\eta)$ and $
		\kappa(\eta)=\int_{\eta}^{\eta_{0}}d\eta a(\eta)\dot{\kappa}(\eta)
		$. \\
		Eq.~(\ref{Bmode1}) shows that the photon-neutrino scattering ($\kappa \not=0$) results in the non-trivial B-mode $\Delta_B^{(S)}$ and modifies the E-mode $\Delta_E^{(S)}$. This value suggest that the Compton scattering by itself can not generate B-mode without taking into account the tensor type of metric perturbations \cite{zh95,uros,Hu}. Therefore, there are three combined power spectra among B-mode, E-mode and T as in
		\begin{figure}
			\includegraphics[scale=1.1]{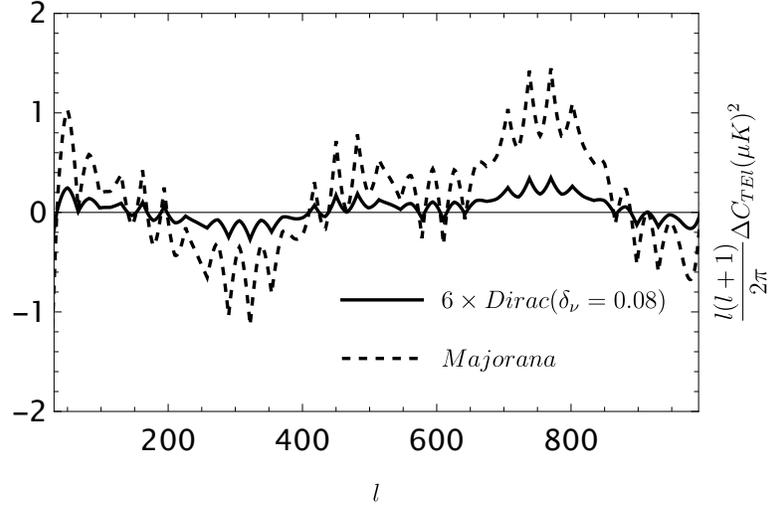}
			\caption{ $\Delta C^{(S)}_{TEl}=C^{(S)}_{TEl}|_{_{CNB}}-C^{(S)}_{TEl}$ via CMB-C$\nu$B forward scattering is plotted in terms of $l$ by considering C$\nu$B as Dirac (solid line) and Majorana (dashed line) neutrino. This plot shows just the contribution of photon-neutrino interaction. Note that corrections from neutrino-photon scattering to TE power spectrum are very smaller than standard cosmology $C^{(S)}_{TEl}$ contribution. Therefore, standard cosmology $C^{(S)}_{TEl}$ contribution is subtracted from the total $C^{(S)}_{TEl}|_{_{CNB}}$ power spectrum in the presence of Thomson and neutrino-photon scattering. We can estimate the contribution of CMB-C$\nu$B interaction on the TE cross power spectrum by $\Delta C^{(S)}_{TEl}/C^{(S)}_{TEl}\simeq 1-\cos(\tilde\kappa)\simeq0.005$ in the case of neutrino as Dirac particles (for $\delta_\nu=0.08$) which may be detected by increasing the precision of future experiments.}
			\label{TE}
		\end{figure}
		\vspace{-1ex}
		\begin{figure*}
			\centering
			\includegraphics[width=0.7\linewidth]{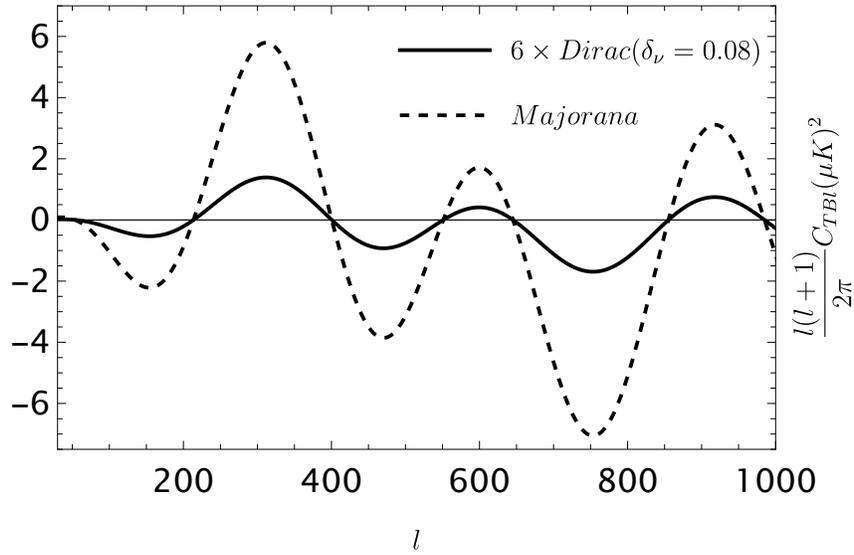}
			
			\caption{TB power spectrum via CMB-C$\nu$B forward scattering is plotted in terms of $l$ by considering C$\nu$B as Dirac (for $\delta_\nu=0.08$)(solid line) and Majorana (dashed line) neutrino. Note that the TB power spectrum in the presence of scalar perturbation is exactly zero in the standard
				scenario of cosmology.}
			\label{TBdata-2}
		\end{figure*}
		{\small 	\begin{eqnarray}
				C_{EB}^{ \ell(S)}&=&(4\pi)^2\frac{9}{16}\frac{(\ell+2)!}{(\ell-2)!}\int K^{2}dK P_S(K)\Delta^{(S)}_{E}\Delta^{(S)}_{B},\label{cross1}
		\end{eqnarray}}
		{\small \begin{eqnarray}
				C_{TE}^{ \ell(S)}&=& (4\pi)^2\sqrt{\frac{9}{16}\frac{(\ell+2)!}{(\ell-2)!}}\int  K^{2}dK P_S(K)
				\Delta^{(S)}_{T}\Delta^{(S)}_{E},\label{cross2}	
		\end{eqnarray}}
		{\small \begin{eqnarray}
				C_{TB}^{ \ell(S)}&=&(4\pi)^2\sqrt{\frac{9}{16}\frac{(\ell+2)!}{(\ell-2)!}}\int K^{2} dK P_S(K) \Delta^{(S)}_T\Delta^{(S)}_{B}, \label{cross3}
		\end{eqnarray}}
		where
		\begin{small}
			\begin{eqnarray}\label{delta}
				&&\Delta^{(S)}_T(K,\eta_0)=\int_0^{\eta_0}d\eta g(\eta)  S_{T}^{(S)}(\eta,K)j_\ell(x),\nonumber\\
				&&\Delta^{(S)}_E(K,\eta_0)=\int_0^{\eta_0}d\eta\,g(\eta)\,\Pi(\eta,K)\frac{j_\ell(x)}{x^2} \cos{\kappa(\eta)},\\
				&&\Delta^{(S)}_B(K,\eta_0)=\int_0^{\eta_0}d\eta\,g(\eta)\,\Pi(\eta,K)\frac{j_\ell(x)}{x^2} \sin{\kappa(\eta)},\nonumber
			\end{eqnarray}
		\end{small}
		and
		\begin{eqnarray}
			&&S_{T}^{^{(S)}}(K,\eta)=g(\Delta_{T0}+2\dot{\alpha}+\dfrac{\dot{v}_{b}}{K}+\dfrac{\Pi}{4}+\dfrac{3\ddot{\Pi}}{4K^{2}})\nonumber\\
			&&+e^{-\tau_{e\gamma}}(\dot{\eta}+\ddot{\alpha})+\dot{g}(\alpha+\dfrac{v_{b}}{K}+\dfrac{3\dot{\Pi}}{4K^{2}})+\dfrac{3\ddot{g}\Pi}{4K^{2}},
		\end{eqnarray}
		where $j_\ell(x)$ is a spherical Bessel function of rank $\ell$, $\dot{\alpha}=(\dot{h}+6\dot{\rho})/2K^2$, $h$, and $\rho$ are additional sources of temperature anisotropy.  The contribution of photon-neutrino (C$\nu$B) scattering to the cross-power spectrum is
		negligible for small $\ell$. However, it is significantly large for $50 < \ell< 200$ ( Figs.(\ref{EBdata-2}-\ref{TBdata-2})).
		As we already know, the non-zero $ C_{TB}^{\ell(S)} $ may also appear in exotic theories due to the presence of helical flows in the primordial plasma at the time of recombination \cite{Pogosian}. Therefore the CMB-C$\nu$B scattering is a new source of TB correlation as well.\\
		The $\kappa$ in Eq.(\ref{delta}) is the effective opacity of neutrino generated by neutrino-photon interaction
		from the last scattering time (redshift $z_l=1100$) to the present time ($z_0=0$). Using the Friedmann equation in the matter-dominated era (in the case of Majorana),
		$H^2/H^2_0=\Omega_M^0(1+z)^{3}+\Omega_\Lambda^0, \, H_0\approx 74\,$km/s/Mpc,
		$\Omega_M^0\approx 0.3\pm0.007, \Omega_\Lambda^0\approx 0.7$ \cite{Planck:2018vyg}, and
		$ad\eta=-dz/H(1+z)$ and conservation of the total neutrino number
		$n_\nu=n_\nu^0 (1+z)^3$, we obtain
		\vspace{-1ex}
		\begin{small}
			\begin{eqnarray}
				\kappa(z) &= &\int^{\eta_{0}}_{\eta}a\,d\eta \dot{\kappa}
				=\frac{\sqrt{2}}{12\pi}\alpha\,G^Fn^0_{\nu}\int^{z}_{0}dz'
				\frac{(1+z')^2}{H(z')}\nonumber\\
				&=&\frac{\sqrt{2}}{12\pi}\alpha\,G^Fn^0_{\nu}\frac{2H(z')}{3\Omega_M^0H_0^{2}}\Big|^{z}_{0}\label{kappa}
				\label{kappabar}
			\end{eqnarray}
		\end{small}
		where the present number-density of all flavor neutrinos and anti-neutrinos is
		$n^0_{\nu}=\sum(n^0_\nu+n^0_{\bar{\nu}})\approx 340\, {\rm cm^{-3}}$.\\
		According to  Eq.(\ref{kappabar}), $\kappa$ strongly depends on the red shift. Thus, it is not wise to ignore this dependency in the integration of Eq.(\ref{delta}). As a result, we can estimate the maximum value $\kappa$ near the last scattering surface with red-shift $z_l=1100$, as follows:
		\begin{eqnarray}
			&&\tilde\kappa= \kappa(z_l)\simeq0.22 (\frac{n_{\nu}^0}{340cm^3})~rad. \label{kappabar1}
		\end{eqnarray}
		If we consider neutrino as Dirac, according to the discussion in  \cite{Geng}, to protect CP conservation, we should use the difference between neutrino and anti-neutrino density. Therefore, we have 
		\begin{eqnarray}
		\kappa(z) &= &\int^{\eta_{0}}_{\eta}a\,d\eta \dot{\kappa}
				=\frac{\sqrt{2}}{12\pi}\alpha\,G^F\delta_\nu n^0_{\nu}\int^{z}_{0}dz'
				\frac{(1+z')^2}{H(z')}\nonumber\\
				&=&\frac{\sqrt{2}}{12\pi}\alpha\,G^F\delta_\nu n^0_{\nu}\frac{2H(z')}{3\Omega_M^0H_0^{2}}\Big|^{z}_{0}
				\label{kappabar1}
		\end{eqnarray}
		where $\delta_\nu$ is the relative density asymmetry of neutrino and anti-neutrino. Now, using the new measurement of the CB angle, we can constrain $\delta_\nu$ as
		\begin{eqnarray}
		\delta_\nu< 0.094 \frac{340 cm^3}{n^0_\nu}.
		\end{eqnarray}
		We can approximately write $ C_{EB}^{l(S)} $, $ C_{TE}^{l(S)} $ and $ C_{TB}^{l(S)} $, for large scale and small $l$ moments, as follows 
		\vspace{-1ex}
		\begin{eqnarray}
			\label{cbestimation}
			&&C_{EB}^{l(S)}\cong \dfrac{1}{2}\sin (2\tilde{\kappa}) \bar{C}_{EE}^{l(S)} \nonumber\\
			&&C_{TE}^{l(S)}\cong \cos (\tilde{\kappa}) \bar{C}_{TE}^{l(S)} \\
			&&C_{TB}^{l(S)}\cong \sin (\tilde{\kappa}) \bar{C}_{TE}^{l(S)}\nonumber
		\end{eqnarray}
		where $\bar{C}_{EE}^{l(S)}$ and $\bar{C}_{TE}^{l(S)}$ are power spectra of E-mode polarization and its correlation with temperature anisotropy  contributed from the Compton scattering alone in the case of scalar perturbation.\\
		\vspace{-2.6ex}
		\begin{figure}
			\includegraphics[scale=0.8]{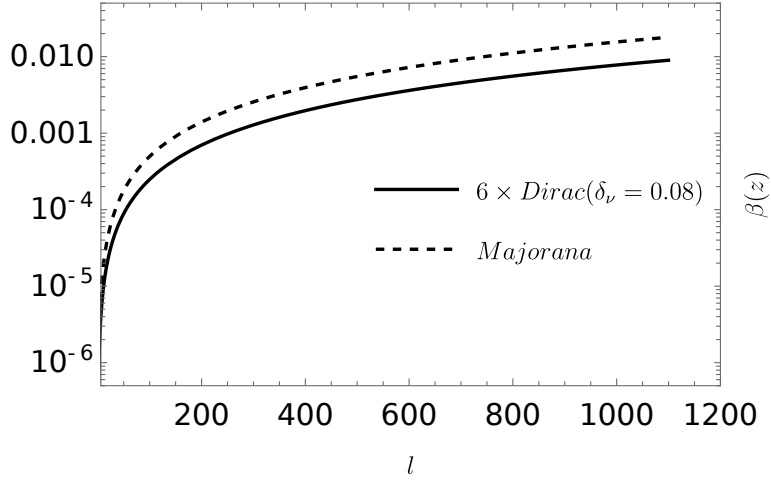}
			\captionof{figure}{ Variation of $\beta=1/2 \kappa$ in radian in terms of redshift.}
			\label{kappa1}
		\end{figure}
			\section{Cosmic birefringence calculation}
			In the presence of CB medium, E-mode can be converted to B-mode and vice versa, so:
			\begin{eqnarray}
				&&\Delta_E^{(S)}=\bar\Delta_E ^{(S)}\cos(2\beta)-\bar\Delta_B^{(S)}\sin(2\beta),\nonumber\\
				&&\Delta_B^{(S)}=\bar\Delta_B^{(S)} \cos(2\beta)+\bar\Delta_E^{(S)}\sin(2\beta),\label{d-beta}
			\end{eqnarray}
			where $\bar\Delta_E ^{(S)}$ and $\bar\Delta_B ^{(S)}$ are the value of E-mode and B-mode, respectively,  in a standard scenario of cosmology without considering CB effects \cite{Minami:2019ruj}. Since $\bar\Delta_B ^{(S)}=0$  in the standard scenario of cosmology, we can rewrite the above equations as
			\begin{eqnarray}
				&&\Delta_E^{(S)}=\bar\Delta_E ^{(S)}\cos(2\beta),\nonumber\\
				&&\Delta_B^{(S)}=\bar\Delta_E^{(S)}\sin(2\beta),\label{d-beta1}
			\end{eqnarray}
			As shown in Eq.(\ref{delta}), we faced an anisotropic birefringence medium. On the other hand,  the $\kappa$ value strongly depends on $\eta$ and red-shift ( Fig.(\ref{kappa1})). The presence of Bessel functions in Eq.(\ref{delta}), which also depends on $\eta$, highlights the importance of $\kappa$ dependence on red-shift. According to this red-shift dependency in the integration of Eq.(\ref{delta}), comparing the theoretical results of $C^{l(S)}_{EB}$ with its reported measurements to extract information is more logical. To give an effective sense,  the CB angle $\beta$ can be estimated by the value of opacity $\kappa(\eta)$ around the last scattering redshift ($z=1100$). The explanation is that the visibility function ($g(\eta)$) in the integrated Eq.(\ref{delta}) has a strong peak at that time. 
			By substituting  $\tilde\kappa$ given in Eq.(\ref{kappabar})  instead of $\kappa$ in Eq.(\ref{delta}) and comparing the results with Eq.(\ref{d-beta1}), we have $\beta\simeq\frac{1}{2} \tilde\kappa$. Finally, we have
			\begin{eqnarray}\label{PS1}
				&&C^{l(S)}_{EE}=\bar{C}^{l(S)}_{EE}\cos^2\tilde\kappa,\\
				&&C^{l(S)}_{BB}=\bar{C}^{l(S)}_{EE}\sin ^2\tilde\kappa,\label{PS2}\\
				&&C^{l(S)}_{EB}=\frac{1}{2}\bar{C}^{l(S)}_{EE}\sin2 \tilde\kappa.\label{PS3}
			\end{eqnarray}
			
			\section{ Big inconsistency between CB angle generated via Majorana C$\nu$B-CMB forward scattering and one reported CB angle}
			
			Note that from the above equations, CB angle $\beta$ can be estimated:
			{\small \begin{equation}\label{alpha}
					\beta|_{\nu\,M}\cong \frac{1}{2} \tilde\kappa=12.6^\circ,
			\end{equation}}
			where $\beta|_{\nu\,M}$ refer to the contribution of C$\nu$B-CMB forward scattering for CB angle in the case of Majorana (M). However, this value is at least one order larger than the cosmic birefringence angle reported by the Planck data release,  $\beta=0.30^\circ\pm0.11^\circ$ ($68\%C.L.$).\par
			As mentioned above, $\kappa$ depends on redshift. Therefore, this point becomes more important, when considering the properties of the Bessel functions $j_\ell(x)$ in Eq.(\ref{delta}). Extracting Eqs.(\ref{PS1}-\ref{PS3}) from Eqs.(\ref{cross1}-\ref{cross3}) can be relevant just in case of ignoring the red-shift dependency of $\kappa$. Here, $\frac{1}{2}\tilde\kappa$ is the highest possible value for CB angle. As a result, to have a more relevant estimation for the CB angle, it is better to directly compare $C^l_{EB}$ generated via C$\nu$B-CMB forward scattering (Fig.\ref{EBdata-2}) by experiment measurements. As Fig.(\ref{EB_EE}) shows, for $l>20$, $C^{l(S)}_{EB}/C^{l(S)}_{EE}\simeq 1/2\sin[2\tilde\kappa]$ , which is compatible with the CB angle given in Eq.(\ref{alpha}).\\
			\begin{figure}
				\includegraphics[scale=1]{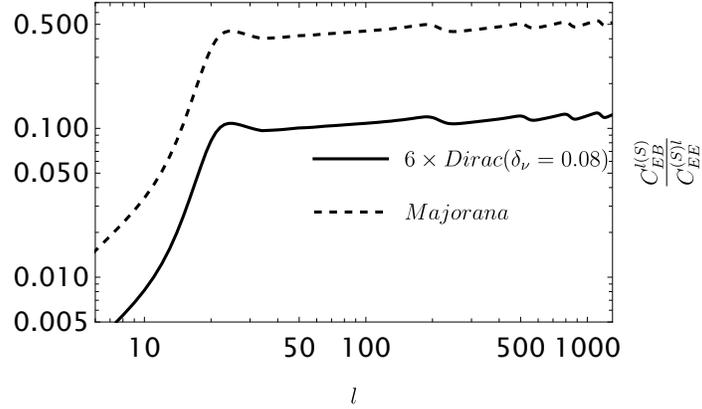}
				\captionof{figure}{ The ratio of $C^{l(S)}_{EB}/C^{l(S)}_{EE}$ is plotted in terms of $l$.}
				\label{EB_EE}
			\end{figure}
			The next point is that here we only calculate the contribution of photon-neutrino forward scattering. There may be some other mechanisms generating the EB power spectrum that can suppress the contribution of photon-neutrino interaction and CB angle. among these mechanisms are Chern-Simons Axion coupling to the electromagnetic field \cite{Gluscevic}, scattering of CMB photons off polarized electrons \cite{Khodagholizadeh:2019het,polarized},  and dipole asymmetry of temperature anisotropy of CMB \cite{dipoleasymmetry}. Large-scale magnetic fields and other Lorentz-violating interactions can also play roles as birefringence mediums. Compton scattering in the presence of non-commutative background and scalar mode of perturbation, in addition to generating circularly polarized microwaves, can lead to B-mode polarization of the Cosmic Microwave Background  \cite{NC} and EB power spectrum. Therefore, it is necessary to check its effect on the CB angle. For example, considering the CB angle due to both C$\nu$B-CMB forward scattering and a Chern-Simons coupling of a time-dependent Axion field to the electromagnetic tensor  and comparing the result with the reported CB angle can provide new constraints on mass and electromagnetic coupling of axion \cite{sadeghobata}.  The role of other cosmological parameters (e.g.,neutrino density, effective flavor number of neutrinos, and Hubble parameter) cannot be ignored for the exact value of the CB angle. Nevertheless, these effects do not change the order of CB angle due to Majorana C$\nu$B-CMB forward scattering.
			\section{ Summary}
			In this study, we wrote quantum Boltzmann equations in the presence of scalar perturbation of metric for temperature fluctuations and linear polarization of CMB radiation by using CMB Stokes parameters (Eq.(\ref{Boltzmann1}). To this end, we considered  CMB-C$\nu$B forward scattering in the context of the standard model of particles as the collision term. 
			Afterward, the explicit expressions for TE, TB, and EB cross-power spectra were derived in the form of the line-of-sight integral solutions.
			We showed that the parity-violating photon-neutrino forward scattering generated non-zero TB and EB power spectra in the presence of scalar perturbation. TE, TB, and EB power spectra via Compton scattering with the effect of CMB-C$\nu$B forward scattering were plotted in terms of $l$ for C$\nu$B as Dirac and Majorana particles  ( Figs.(\ref{EBdata-2}-\ref{TBdata-2})). As a result, TB and EB power spectra in terms of $l$ behaved like  TE and EE power spectra, respectively. \\
			Also, it was proved that  C$\nu$B-CMB forward scattering can make an opaque medium that can play as a birefringence medium and rotates the linear polarization plane of CMB. As shown in Eq.(\ref{delta}),  this medium is an anisotropic birefringence medium and $\kappa$ value depends on the red-shift Fig.(\ref{kappa1}). The presence of Bessel functions (and also dependence on $\eta$) in Eq.(\ref{delta}) highlights the importance of $\kappa$ dependence on red-shift. In such a case, comparing $C^{l(S)}_{(EB)}$ with its reported measurements is the most reasonable decision at least in the case of $l<20$ (Fig.(\ref{EB_EE})). Our results for $C^{l(S)}_{(EB)}$ are shown in Fig.(\ref{EBdata-2}).\\
			Despite the extreme dependence of $\kappa$ on the red shift, to give an effective sense, we showed that the CB angle $\beta$ can be roughly estimated by the value of opacity $\beta|_\nu\simeq1/2\tilde\kappa$ in the case of $l>20$ (Fig.(\ref{EB_EE})).  Therefore, we obtained $\beta|_{\nu\,M}\simeq12.6^\circ$, which refers to the contribution of C$\nu$B-CMB forward scattering for CB angle in the case of C$\nu$B as Majorana particles. It seems this result is not in agreement with the measurement reported in \cite{minami}.
			Also, consistent with\cite{Geng}, we showed that if the nature of C$\nu$B is Dirac, the last reported measurement of CB angle from Planck data release ($\beta=0.3^\circ \pm 0.11^\circ$) can constrain relative density asymmetry of neutrino and anti-neutrino by the $\delta_\nu<0.094$.
			\par
			The next point is that we only calculated the contribution of C$\nu$B-CMB forward scattering. In this respect, some other mechanisms can reduce the contribution of  C$\nu$B-CMB interaction and CB angle. Some of these mechanisms are Axion-CMB interaction, polarized Compton scattering, Lorentz violating interaction, external large-scale magnetic field, and dipole asymmetry of CMB temperature anisotropy, that generate EB power spectrum. Furthermore, the role of other cosmological parameters cannot be ignored on the exact value of CB angle, e.g., neutrino density and effective flavor number of neutrinos.  Hopefully, by upgrading data of EB-cross power spectrum and, theoretical and experimental studies in the future, we can obtain more information about the origin of the reported CB angle, the nature of C$\nu$B and Axion mass, and its electromagnetic coupling.
			\section{Appendix}
			\subsection{Neutrino Calculation}
			Here, we show more details of Eq.(\ref{QBE}) for neutrinos. Interaction Lagrangian is
			\begin{eqnarray}\label{lagarangian}
				\mathcal{L}_I=\mathcal{L}_{QED}+\mathcal{L}_{e\nu}
			\end{eqnarray}
			where the first term is QED Lagrangian and the second term is weak interaction containing
			electron-neutrino-W boson vertex. The Photon field and neutrino field are considered as follows,
			\begin{eqnarray}
				&&	A_\mu(x) = \int \frac{d^3 k}{(2\pi)^3
					2 k^0} \left[ a_i(k) \epsilon _{i\mu}(k)
				e^{-ik\cdot x}+ a_i^\dagger (k) \epsilon^* _{i\mu}(k)e^{ik\cdot x}
				\right]\\
				&&	\psi(x) = \int \frac{d^3 q}{(2\pi)^3}\frac{1}{
					\sqrt{2 E_{\mathbf{q}}}} \sum_r\left[ b_r(p) \mathcal{U}_{r}(p)
				e^{-iq\cdot x}+ d_r^\dagger (q) \mathcal{V}_{r}(q)e^{iq\cdot x}
				\right],\label{psi}
			\end{eqnarray}  
			where $\epsilon_{i\mu}$ is the polarization vector of the photon, $\mathcal{U} and \mathcal{V}$ are Dirac Spinors and $a,a^\dagger$, $b,b^\dagger$ and $d,d^\dagger$ are annihilation and creation operators of photon, neutrino, and anti-neutrino respectively with the following commutator relations,
			\begin{eqnarray}\label{commutator}
				&&[a_s(k),a^\dagger_{s'}(k')]=(2\pi)^32k^0\delta_{ss'}\delta^{(3)}(k-k')~,\nonumber\\
				&&\{b_r(q),b^\dagger_{r'}(q')\}=\{d_r(q),d^\dagger_{r'}(q')\}
				=(2\pi)^3\delta_{rr'}\delta^{(3)}(q-q')	
			\end{eqnarray}
			In the SM for elementary particle physics, the Hamiltonian of photon-neutrino forward scattering at the leading order is
			given by the exchange of leptons and the $W-Z$ gauge bosons, via the one-loop diagrams, as shown in Fig.(7),
			\bea
			H_{I}^0&=&\int d\mathbf{q}d\mathbf{q}'d\mathbf{k}d\mathbf{ k}'(2\pi)^3\delta^3(\mathbf{ q}'-\mathbf{ q})(2\pi)^3\delta^3(\mathbf{ k}'-\mathbf{ k})\exp\left[i\left(q'^0+p'^0-q^0-p^0\right)t\right]
			\nonumber\\ && \times
			\left[ b^{\dagger}_{r'}(\mathbf{ q}')a^{\dagger}_{s'}(\mathbf{ p}')T_{fi}(ps,qr,p's',q'r')b_{r}(\mathbf{ q})a_{s}(\mathbf{ p})\right]~.\label{h0}
			\eea
			where neutrino phase spaces $\int d\mathbf{q}\equiv d^3q/\sqrt{2q^0}$ and $\int d\mathbf{q}'\equiv d^3q'/\sqrt{2q'^0}$.
			The scattering amplitude $T_{fi}$ is given by
			\begin{eqnarray}
				T_{fi} &=& \frac{1}{8}e^2g_w^2\int\frac{d^4l}{(2\pi)^4}D_{\alpha\beta}(q-l)\bar{\mathcal{U}}_{r'}(\mathbf{q}')\gamma^\alpha (1-\gamma_5)S_F(l+p-p')\nonumber\\
				&\times& \left[\ep_{s'}S_F(l+p)\ep_{s}+\ep_{s}S_F(l-p')\ep_{s'}\right]
				S_F(l)\gamma^\beta (1-\gamma_5)\mathcal{U}_r(\mathbf{q})~,\label{md}
			\end{eqnarray}
			where the notations
			$D_{\alpha\beta}$ and $S_F$ represent the propagators of $W_\mu^\pm$ gauge-bosons and charged-lepton respectively. By helping Dimensional regularization and Feynman parameters, we go forward to obtain  the leading order terms of the right side of the above equation. Then
			\begin{eqnarray}
				T_{fi} &=& -\frac{1}{16}\frac{1}{4\pi^2}e^2g_w^2\,
				\int_0^1dy\int_0^{1-y}dz\,\frac{(1-y-z)}{zM^2_W}\,\bar{\mathcal{U}}_{r}(q)(1+\gamma_5)\nonumber\\&\times&\Big(2z q\!\!\!/\epsilon_{s'}.\epsilon_s
				+ \,2z(\ep_{s'}\,\,\mathbf{q}.\epsilon_s\,+\ep_{s}\,\mathbf{q}.\epsilon_{s'}\,)+
				(3y-1)k\!\!\!/\,(\ep_{s}\,\ep_{s'}\,-\ep_{s'}\,\ep_{s}\,)\,\Big)
				\mathcal{U}_r(q).\label{fws2}
			\end{eqnarray}
			Here we use the gamma-matrix identity $A\!\!\!\!/\,B\!\!\!\!/=2A.B-B\!\!\!\!/\,A\!\!\!\!/$, the polarization vector properties $k.\epsilon_i=0$ and
			$\epsilon_i.\epsilon_j=-\delta_{ij}$. The straightforward calculations of Eqs.~(\ref{h0}) and (\ref{md}) yield
			the one-loop effective Hamiltonian of the photon-neutrino forward scattering  \cite{Mohammadi:2013ksa}
			\begin{eqnarray}
				H_{I}^{0} &=& \alpha\,G^F\frac{\sqrt{2}}{12\pi}\,\int\frac{d^3k}{(2\pi)^3}\,\left[a^{\dagger}_{s'}(\mathbf{k})\,a_s(\mathbf{k})\right] \,\int \frac{d^3q}{(2\pi)^3}\,b^{\dagger}_{r}(\mathbf{q})\,b_r(\mathbf{q}) \nonumber \\
				&&\times\Big( 2\epsilon_{s'}(\mathbf{p})\cdot\epsilon_s(\mathbf{p})(q^0-|\mathbf{q}|)
				+4|{\bf q}|(\hat{\mathbf{q}}\cdot\epsilon_s)\,(\hat{\mathbf{q}}\cdot\epsilon_{s'})
				-2i\varepsilon_{\lambda\alpha\beta\mu}k^\alpha\,\epsilon_{s'}^\beta\,\epsilon_s^\mu\,\hat{q}^{\lambda} \Big).
				\label{h2}
			\end{eqnarray}
			Now, using the above interaction Hamiltonian, 
			commutators (\ref{commutator}) and below expectation values of operators are obtained
			\begin{eqnarray}
				\langle \, a_1a_2...b_1b_2...\, \rangle
				&=&\langle \,a_1a_2...\, \rangle\langle \,b_1b_2...\, \rangle,
				\label{contraction1}\\
				\langle \, a^\dag_{s'}(k')a_{s}(k)\, \rangle
				&=&2k^0(2\pi)^3\delta^3(\mathbf{k}-\mathbf{k'})\rho_{ss'}(\mathbf{x},\mathbf{k}),
				\label{contraction2}\\
				\langle \, b^\dag_{r'}(q')b_{r}(q)\, \rangle
				&=&(2\pi)^3\delta^3(\mathbf{q}-\mathbf{q'})\delta_{rr'}\frac{1}{2}n_\nu(\mathbf{x},\mathbf{q}),
				\label{contraction3}\\
				\langle \, d^\dag_{r'}(q')d_{r}(q)\, \rangle
				&=&(2\pi)^3\delta^3(\mathbf{q}-\mathbf{q'})\delta_{rr'}\frac{1}{2}n_{\bar\nu}(\mathbf{x},\mathbf{q}),
				\label{contraction4}
			\end{eqnarray} 
			where $\rho_{ss'}(\mathbf{x},\mathbf{k})$ is the local matrix density and $n_\nu(\mathbf{x},\mathbf{q})$ and $n_{\bar\nu}(\mathbf{x},\mathbf{q})$ are the local spatial density of neutrinos and anti-neutrinos in the momentum state $\mathbf{q}$, respectively.  We can calculate the forward scattering term of Eq.(\ref{QBE}). 
			After tedious but straightforward calculations, we have
			\begin{eqnarray}
				&&(2\pi)^32k^0\frac{d\rho_{ij}}{dt}=i\langle[H^0_I, D^0_{ij}({\bf k})]\rangle = \frac{-e^2g_w^2}{64\pi^2M_W^2}\int d\mathbf{q} \big[\rho_{s'j}(\mathbf{x},{\bf k})\delta_{is} -\rho_{is}(\mathbf{x},{\bf k})\delta_{js'}\big]n_\nu(\mathbf{x},\mathbf{q})\nonumber\\
				&\times& \bar{\mathcal{U}}_{r}(q)(1+\gamma_5)\Big[ q\!\!\!/\epsilon_{s'}\cdot\epsilon_s
				+ \,(\ep_{s'}\,\,q\cdot\epsilon_s\,+\ep_{s}\,q\cdot\epsilon_{s'}\,)
				+
				\frac{1}{6}k\!\!\!/\,(\ep_{s}\,\ep_{s'}\,-\ep_{s'}\,\ep_{s}\,)\Big]\,\mathcal{U}_r(q)+{O}(\frac{m_e^2}{M_W^4}).
				\label{fws1}
			\end{eqnarray}
			The next step in the Quantum Boltzmann equation solution is to solve Eq.(\ref{Boltzmann}).
			\begin{figure}[H]
				\begin{center}
					\begin{tikzpicture}[scale=0.90,
						thick,
						level/.style={level distance=3.15cm, line width=0.4mm},
						level 2/.style={sibling angle=60},
						level 3/.style={sibling angle=60},
						level 4/.style={level distance=1.4cm, sibling angle=60}
						]
						\node[draw=none,fill=none] (p1) at (-4,3){}  ;
						\node[draw=none,fill=none] (v1) at (-2,2){}  ;
						\node[draw=none,fill=none] (v2) at (0,2){}  ;
						\node[draw=none,fill=none] (v3) at (-2,0.0){}  ;
						\node[draw=none,fill=none] (v4) at (0,0){}  ;
						\node[draw=none,fill=none] (p2) at (-4,-1.5){}  ;
						\node[draw=none,fill=none] (k1) at (1.7,3) {};
						\node[draw=none,fill=none] (k2) at (1.7,-1.5) {};
						\node[draw=none,fill=none] at (-3,3){$k$}  ;
						\node[draw=none,fill=none] at (-1,-.5){$q-p$};
						\node[draw=none,fill=none] at (-1,2.5){$p+k$};
						\node[draw=none,fill=none] at (1,1){$p+k+k'$};
						\node[draw=none,fill=none] at (-2.5,1){$p$};
						\node[draw=none,fill=none] at (-3.1,-0.5){$q$}  ;
						\node[draw=none,fill=none] at (0.5,3) {$ k' $};
						\node[draw=none,fill=none] at (1.1,-.5) {$ q'$};
						\draw[photon]   (p1) -- (v1);
						\draw[electron] (v1) -- (v2);
						\draw[photon]   (v2) -- (k1);
						\draw[electron] (v4) -- (v3);
						\draw[electron] (v3) -- (v1);
						\draw[electron] (v2) -- (v4);
						\draw[electron] (v4) -- (k2);
						\draw[electron] (p2) -- (v3);
						
						\draw[black]         (-2,2) circle (0.1cm);
						\fill[black]    (0,0) circle (0.1cm);
						\fill[black]    (-2,0) circle (0.1cm);
						\draw[black]         (0,2) circle (0.1cm);
					\end{tikzpicture}
					\begin{tikzpicture}[scale=0.90,
						thick,
						level/.style={level distance=3.15cm, line width=0.4mm},
						level 2/.style={sibling angle=60},
						level 3/.style={sibling angle=60},
						level 4/.style={level distance=1.4cm, sibling angle=60}
						]
						\node[draw=none,fill=none] (p1) at (-4,3){}  ;
						\node[draw=none,fill=none] (v1) at (-2,2){}  ;
						\node[draw=none,fill=none] (v2) at (0,2){}  ;
						\node[draw=none,fill=none] (v3) at (-2,0.0){}  ;
						\node[draw=none,fill=none] (v4) at (0,0){}  ;
						\node[draw=none,fill=none] (p2) at (-4,-1.5){}  ;
						\node[draw=none,fill=none] (k1) at (1.7,3) {};
						\node[draw=none,fill=none] (k2) at (1.7,-1.5) {};
						\node[draw=none,fill=none] at (-3,3){$k$}  ;
						\node[draw=none,fill=none] at (-1,-.5){$q-p$};
						\node[draw=none,fill=none] at (-1,1.7){$p-k'$};
						\node[draw=none,fill=none] at (1,1){$p+k+k'$};
						\node[draw=none,fill=none] at (-2.5,1){$p$};
						\node[draw=none,fill=none] at (-3.1,-0.5){$q$}  ;
						\node[draw=none,fill=none] at (0.5,3) {$ k' $};
						\node[draw=none,fill=none] at (1.1,-.5) {$ q'$};
						\draw[photon]   (p1) -- (v2);
						\draw[electron] (v1) -- (v2);
						\draw[photon]   (v1) -- (k1);
						\draw[electron] (v4) -- (v3);
						\draw[electron] (v3) -- (v1);
						\draw[electron] (v2) -- (v4);
						\draw[electron] (v4) -- (k2);
						\draw[electron] (p2) -- (v3);
						
						\draw[black]         (-2,2) circle (0.1cm);
						\fill[black]    (0,0) circle (0.1cm);
						\fill[black]    (-2,0) circle (0.1cm);
						\draw[black]         (0,2) circle (0.1cm);
					\end{tikzpicture}
					\caption{Full detail of the most contributing photon-neutrino Feynman diagrams in $H_I^0(t)$.}
					\captionsetup[figure]{justification=justified,singlelinecheck=true}
					\label{fig:photon-neutrino1}
				\end{center}
			\end{figure}
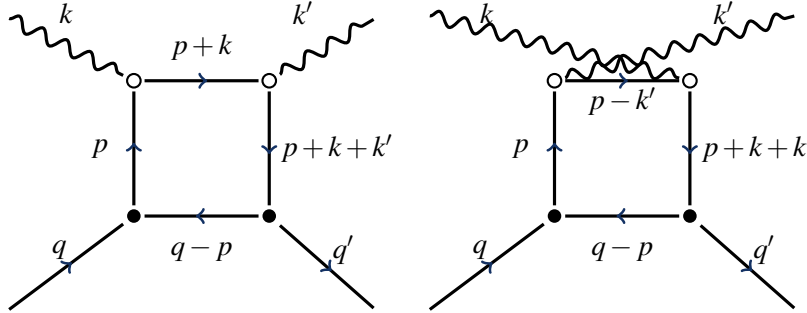
			\subsection{Photon-Neutrino Effective Lagrangian in  six dimensions D=6}
			To give a more convincing, we expand our calculation to anti-neutrino-photon scattering and Majorana neutrino, we have the following gauge invariant effective Lagrangian in 6 dimensions (D=6) \cite{Geng}
			\begin{eqnarray}
				&&	\mathcal{L}_{\gamma\nu}\sim G_F\alpha \epsilon^{\mu\nu\alpha\beta}F_{\mu\nu}\bar\nu_R\gamma_\alpha \nu_L A_\beta. \label{eff2},
			\end{eqnarray}
			This effective Lagrangian for Majorna fermion can be written as 
			\begin{eqnarray}
				&&	\mathcal{L}_{\gamma\nu}\sim G_F\alpha \epsilon^{\mu\nu\alpha\beta}F_{\mu\nu}\chi^{\dagger}\bar{\sigma}_\alpha \chi A_\beta \label{eff3},
			\end{eqnarray}
			where $\chi$ is the massless 2-component of Majorana neutrino. According to the \cite{Geng} paper, the final result of the Boltzmann equation solving Dirac neutrino must be calculated using relative density asymmetry of neutrino and anti-neutrino ($\delta_\nu$) to preserve CP conserving of Eq.({\ref{eff2}}). However, since Eq.({\ref{eff3}}) is completely same for neutrino and anti-neutrino in the case of Majorana neutrino, we should add up neutrino density and anti-neutrino density in the final result.\\
			The above effective Lagrangians contain CP conservation terms. Due to the CP conservation of the above interaction, antineutrino-photon forward scattering in the Dirac case, can generate a CB angle in the opposite sign and equal amplitude in comparison with neutrino-photon forward scattering,
			\begin{eqnarray}
				&&(2\pi)^32k^0\frac{d\rho_{ij}}{dt}=i\langle[H^0_I, D^0_{ij}({\bf k})]\rangle =-\frac{e^2g_w^2}{64\pi^2M_W^2}\int d\mathbf{q} \big[\rho_{s'j}(\mathbf{x},{\bf k})\delta_{is} -\rho_{is}(\mathbf{x},{\bf k})\delta_{js'}\big]n_{\bar{\nu}}(\mathbf{x},\mathbf{q})\nonumber\\
				&\times& \bar{\mathcal{V}}_{r}(q)(1+\gamma_5)\Big[ q\!\!\!/\epsilon_{s'}\cdot\epsilon_s
				+ \,(\ep_{s'}\,\,q\cdot\epsilon_s\,+\ep_{s}\,q\cdot\epsilon_{s'}\,)
				+
				\frac{1}{6}k\!\!\!/\,(\ep_{s}\,\ep_{s'}\,-\ep_{s'}\,\ep_{s}\,)  \Big]\,\mathcal{V}_r(q)+{O}(\frac{m_e^2}{M_W^4}).
				\label{fws3}
			\end{eqnarray}
			and for Majorana, we have
			\begin{eqnarray}
				&&(2\pi)^32k^0\frac{d\rho_{ij}}{dt}=i\langle[H^0_I, D^0_{ij}({\bf k})]\rangle =-\frac{e^2g_w^2}{64\pi^2M_W^2}\int d\mathbf{q} \big[\rho_{s'j}(\mathbf{x},{\bf k})\delta_{is} -\rho_{is}(\mathbf{x},{\bf k})\delta_{js'}\big]n_{\bar{\nu}}(\mathbf{x},\mathbf{q})\nonumber\\
				&\times& \chi^\dagger_r(q)(1+\gamma_5)\Big[ q\!\!\!/\epsilon_{s'}\cdot\epsilon_s
				+ \,(\ep_{s'}\,\,q\cdot\epsilon_s\,+\ep_{s}\,q\cdot\epsilon_{s'}\,)
				+
				\frac{1}{6}k\!\!\!/\,(\ep_{s}\,\ep_{s'}\,-\ep_{s'}\,\ep_{s}\,)  \Big]\,\chi_r(q)+{O}(\frac{m_e^2}{M_W^4}).
				\label{fws4}
			\end{eqnarray}

			As a result, in Eqs.(\ref{kappabar} and \ref{kappabar1}), we should consider $n^0_{\nu}=\sum(n^0_\nu+n^0_{\bar{\nu}})\approx 340\, {\rm cm^{-3}}$. \par
			In some references such as  \cite{Karl:2004bt}, authors discussed various parity-odd and parity-even effective Lagrangians that describe photon-neutrino interaction. As  can be seen, all these effective Lagrangians are dimension-8 operators or higher ones with  $\alpha^2G_F^2$ order. As a result, their calculations of one-loop scattering amplitude are several orders of magnitude smaller than our results ($\alpha\, G_F$). In addition,  in a recently published work \cite{Dvornikov:2020olb}, authors calculated the CB angle in presence of relic neutrino-antineutrino gas using electroweak correction of vacuum polarization of the photon. however, this effect is proportional to $\frac{\alpha}{M_W^4}$, which is very smaller than our results.
			
		\end{document}